\documentclass[prl,twocolumn]{revtex4}

\begin{document}
\title{Two Different Definitions on Genuine Quantum and Classical Correlations
in Multipartite Systems Do Not Coincide in General}

\author{Zhanjun Zhang$^1$}
\author{Biaoliang Ye$^1$}
\author{Shao-Ming Fei$^{2,3}$}
\affiliation{$^1$School of Physics \& Material Science, Anhui
University, Hefei 230039, China\\
$^2$Department of Mathematics, Capital Normal University, Beijing
100037, China\\
$^3$Max-Planck-Institute for Mathematics in the Sciences, 04103
Leipzig, Germany}

\maketitle

Giorgi et al \cite{1} presented an approach to study genuine quantum and classical correlations in multipartite systems. They defined total information $T(\varrho)$, total classical correlation
${\cal J}({\varrho})$, and total quantum discord ${\cal D}(\varrho)= T(\varrho)-{\cal J}(\varrho)$
for a tripartite state $\varrho$, and the genuine tripartite correlation $T^{(3)}(\varrho)=T(\varrho)-T^{(2)}(\varrho)$, where
$T^{(2)}(\varrho)={\rm max}[{\cal I}(\varrho_{a,b}),{\cal I}(\varrho_{a,c}),{\cal I}(\varrho_{b,c})]$
with ${\cal I}$ being the bipartite mutual information in $\varrho$.
Without loss of generality, suppose $T^{(3)}(\varrho)=T(\varrho)-{\cal I}(\varrho_{a,b})$.
Then they defined the concerned genuine tripartite classical and quantum correlations,
i.e., ${\cal J}^{(3)}(\varrho)$ and ${\cal D}^{(3)}(\varrho)$, in two different ways (labeled by single and double primes respectively):
1) $T^{(3)}(\varrho)={\cal J}^{(3)'}(\varrho)+{\cal D}^{(3)'}(\varrho)$,
where $T^{(3)}(\varrho)$ is the bipartite mutual information between subsystems $ab$ and $c$;
and 2)
\begin{equation}\label{1}
\begin{array}{rcl}
{\cal J}^{(3)''}(\varrho)&=&{\cal J}(\varrho)-{\cal J}^{(2)}(\varrho),\\
{\cal D}^{(3)''}(\varrho)&=&{\cal D}(\varrho)-{\cal D}^{(2)}(\varrho),
\end{array}
\end{equation}
where ${\cal J}^{(2)}(\varrho)={\rm max}[{\cal J}(\varrho_{a,b}),{\cal J}(\varrho_{a,c}),{\cal J}(\varrho_{b,c})]$
and ${\cal D}^{(2)}(\varrho)={\bf max}[{\cal D}(\varrho_{a,b}),{\cal D}(\varrho_{a,c}),{\cal D}(\varrho_{b,c})]$.
The authors ``must show that the two definitions coincide". However they
specialized on the case of three-qubit pure states and shew the coincidence in this case.
Nonetheless, we find that the two definitions do not coincide in general.

We first remark that ${\cal D}^{(2)}(\varrho)$
in Ref.\cite{1} is mistakenly written as ${\cal D}^{(2)}(\varrho)={\bf min}[{\cal D}(\varrho_{a,b}),
{\cal D}(\varrho_{a,c}),{\cal D}(\varrho_{b,c})]$. According to the sentence above
the equation (4) in Ref.\cite{1}, the minimization should be maximization. Otherwise,
the two definitions disagree even for three-qubit pure states. Because for a three-qubit pure state $\varrho$,
${\cal D}(\varrho_{a,b})={\cal I}(\varrho_{a,b})-{\cal J}(\varrho_{a,b})
=S(\varrho_a)-S(\varrho_c)+{\cal E}(\varrho_{b,c})$ and
${\cal D}(\varrho_{a,b}) \geq {\rm max}[{\cal D}(\varrho_{a,c}),{\cal D}(\varrho_{b,c})]$
in the case of ${\cal I}(\varrho_{a,b}) \geq {\cal I}(\varrho_{a,c})\geq {\cal I}(\varrho_{b,c})$.
Hence, ${\cal D}(\varrho_{a,b})$ is not the minimal bipartite quantum discord.
Alternatively, the second equality of equation (8) in Ref.\cite{1} does not hold, i.e.,
${\cal D}^{(2)}(\varrho) \neq S(\varrho_a)-S(\varrho_c)+{\cal E}(\varrho_{b,c})$.
As a result, the equation (9) in Ref.\cite{1} does not hold according to the second definition,
but does hold according to the first definition.

Let us now consider a three-qubit mixed state $\varrho=\frac12[(|0\rangle\langle0|)_b
\otimes \varrho_{a,c}(0.1,3\pi/10)+ (|1\rangle\langle1|)_b \otimes \varrho_{a,c}(0.7,\pi/5)]$,
where $\varrho_{a,c} (p,\theta)= [(1-p)|00\rangle \langle00|+ p\sin^2\theta |01\rangle \langle01|
+ p\sin\theta\cos\theta |01\rangle \langle10| + p\sin\theta\cos\theta|10\rangle \langle01|
+ p\cos^2\theta |10\rangle \langle10|]_{a,c}$. One has
${\cal I}(\varrho_{a,b})\approx 0.27$, ${\cal I}(\varrho_{a,c})\approx 0.22$,
${\cal I}(\varrho_{b,c})\approx 0.01$. Hence, $T^{(2)}(\varrho)={\cal I}(\varrho_{a,b})$.
Note that $\varrho$ is actually a bipartite classical-quantum correlated state
in the bipartite partition $b$ and $ac$. Consequently, in terms of Refs. \cite{2,3}
one can get ${\cal D}(\varrho_{a,b})={\cal D}(\varrho_{b,c})=0$, i.e.,
${\cal J}(\varrho_{a,b})={\cal I}(\varrho_{a,b})$ and ${\cal J}(\varrho_{b,c})={\cal I}(\varrho_{b,c})$.
Moreover, $\varrho_{a,c}$ is actually a bipartite entangled state, since the entanglement of formation \cite{4} of $\varrho_{a,c}$ is positive, ${\cal E}(\varrho_{a,c})\approx 0.11 >0$.
Consequently, ${\cal D}(\varrho_{a,c})>0$ and ${\cal J}(\varrho_{a,c})< {\cal I}(\varrho_{a,c})$.
Therefore ${\cal J}^{(2)}(\varrho)$ is just ${\cal J}(\varrho_{a,b})$ and ${\cal D}^{(2)}(\varrho)$ is ${\cal D}(\varrho_{a,c})$. At last we have ${\cal J}^{(3)''}(\varrho) + {\cal D}^{(3)''}(\varrho)=
[{\cal J}(\varrho)-{\cal J}^{(2)}(\varrho)]+[{\cal D}(\varrho)-{\cal D}^{(2)}(\varrho)]
=T(\varrho)-{\cal J}^{(2)}(\varrho)-{\cal D}(\varrho_{a,c})
=T^{(3)}(\varrho)- {\cal D}(\varrho_{a,c})$.
In contrast, we get ${\cal J}^{(3)'}(\varrho) + {\cal D}^{(3)'}(\varrho)=T^{(3)}(\varrho)$.
Obviously, both definitions on ${\cal J}^{(3)}(\varrho)$ and
${\cal D}^{(3)}(\varrho)$ disagree for this simple example.

Generalizing to the multipartite case, the authors further stated that
for $n$-partite pure states it is still possible to define, through a ladder procedure,
${\cal D}^{(n)}$ and ${\cal J}^{(n)}$ in analogy with Eq.(\ref{1}).
Since the coincidence can not be completely assured
for all tripartite states, the coincidence of the two definitions after the generalization via the so-called
ladder procedure to multipartite cases is again not reliable. Incidentally, for
an $n$ $(n\geq 4)$-partite pure state, its reduced tripartite states are mixed ones in general.
For instance, as the reduced state of the 6-qubit pure state $|\Psi\rangle_{abca'b'c'}= \{ |00\rangle_{bb'} [\sqrt{0.9}|00\rangle_{ac}|00\rangle_{a'c'} + \sqrt{0.1}(\sin\frac{3\pi}{10}|01\rangle_{ac}
+\cos\frac{3\pi}{10}|10\rangle_{ac}) |01\rangle_{a'c'}] + |11\rangle_{bb'}[\sqrt{0.7}|00\rangle_{ac}|10\rangle_{a'c'}
+\sqrt{0.3}(\sin\frac{\pi}{5}|01\rangle_{ac}+\cos\frac{\pi}{5}|10\rangle_{ac}) |11\rangle_{a'c'}]
\}/\sqrt 2$, $\rho_{abc}={\rm Tr}_{a'b'c'}(|\Psi\rangle\langle \Psi|)_{abca'b'c'}$ is just the tripartite mixed state in our example above in revealing the disagreement.

In summary, the genuine quantum and classical correlations defined by Giorgi et al \cite{1}
in two different ways do not coincide in general.

\medskip
Work supported by the Specialized Research Fund for the Doctoral Program of
Higher Education under Grant No. 20103401110007, the NNSFC under Grant Nos. 10975001,
the 211 Project of Anhui University, and PHR201007107.

\end{document}